# The effect of website attributes and mental involvement online impulse purchases

Seyedeh Samira Saadatmand[*]

**Abstract:**
This research aimed at investigating the impact of website features and involvement on immediate online shopping. The research is applied in terms of type and it is causative in terms of methodology. The statistical population consisted of all citizens of Tabriz, who have purchased clothes online at least once and 260 individuals were chosen randomly and the questionnaires were collected according to this sample size. The data were collected by questionnaire. For analysis of the data, software "SPSS" and for test of the model hypotheses, SEM was used by confirmatory factor analysis. The results showed that the website benefit-oriented features have a positive impact on immediate online shopping and website benefit-oriented features have no significant impact on immediate online shopping.

---

[*] M.A. Student of Business Management, International Business, ALGHADIR Non- Governmental and Private Higher Education Institution, Tabriz, Iran

# بررسی تاثیر ویژگی‌های وب سایت و درگیری ذهنی بر خریدهای آنی اینترنتی

**سیده سمیرا سعادتمند** *

دانشجوی کارشناسی ارشد رشته مدیریت بازرگانی گرایش بازرگانی بین الملل، موسسه آموزش عالی الغدیر، تبریز، ایران

**چکیده:**

هدف پژوهش حاضر بررسی تاثیر ویژگی های وب سایت و درگیری ذهنی بر خریدهای آنی اینترنتی می باشد. این پژوهش از لحاظ هدف از نوع تحقیقات کاربردی و از حیث روش از نوع تحقیقات علی می باشد. جامعه آماری پژوهش شامل کلیه شهروندان تبریزی می باشد، که حداقل یکبار از طریق اینترنت پوشاک خریداری کرده اند، ۲۶۰ نمونه به صورت تصادفی انتخاب پرسش نامه‌ها جمع آوری گردید. ابزار گردآوری اطلاعات در این پژوهش پرسشنامه می باشد. برای تحلیل اولیه داده ها از نرم افزار «SPSS» و برای آزمون فرضیه‌های مدل، از مدلسازی معادلات ساختاری «SEM»، با بهره‌گیری از تحلیل عاملی تاییدی استفاده شده است. نتایج پژوهش آشکار کرد که درگیری ذهنی بر خرید آنی اینترنتی تاثیر مثبت دارد. همچنین نتایج پژوهش نشان داد که ویژگی‌های نفع‌گرایانه وب سایت بر خرید آتی اینترنتی تاثیر مثبت دارد ولی ویژگی‌های لذت‌گرایانه وب سایت بر خرید آنی اینترنتی تاثیر معنادار ندارد.

**واژگان کلیدی:** ویژگیهای وب سایت- درگیری ذهنی- خرید آنی اینترنتی

---

\* نویسنده مسئول: samirasaadatmand58@gmail.com

## مقدمه:

## طرح مسئله:

با گسترش فناوری اطلاعات در جهان و ورود سریع آن به زندگی روزمره، مسایل و ضرورت های تازه‌ای بوجود آمده است و کسب و کار الکترونیکی جایگزین روشهای سنتی شده است. با آشکار ساختن اهمیت نسبی عوامل موثر بر رفتار خرید آنی، می‌توان استراتژی‌های بازاریابی اثربخش پیشنهاد داد تا حجم خرید آنی یک فروشگاه اینترنتی افزایش یابد و از سوی دیگر می‌توان به مصرف کنندگان کمک کرد تا رفتار خرید آنی خود را کنترل نمایند. آگاهی از عوامل تاثیر گذار بر خرید اینترنتی مصرف کنندگان و خریداران برای مدیران فروشگاههای اینترنتی امری ضروری است و به آنها کمک می‌کند تا با برنامه ریزی بهتر، سایت و فروشگاه اینترنتی خود را متناسب با نیاز استفاده کنندگان طراحی کند. خرید آنی اینترنتی بخش اعظمی را شامل می‌شود که از منطق پیروی نمی‌کنند و احساسی هستند. مطالعات متعددی نشان داده است، افرادی که دارای هریک از درگیری ذهنی بالا (زیاد) و پایین (کم) هستند واکنش های مختلفی به خرید آنی اینترنتی دارند. همچنین مصرف کنندگان با ویژگی‌های منفعت‌گرایانه می‌بایست از طریق استدلال‌های منطقی قانع شوند، تا یک خریدار واقعی باشند، مصرف‌کنندگان با ویژگی‌های لذت‌گرایانه از طریق عواطف و محرک-های احساسی به یک خریدار واقعی تبدیل می‌شوند. بنابراین برای یک خرید آنی اینترنتی درگیری ذهنی و ویژگی‌های وب سایت می‌تواند تاثیرگذار باشد و به عنوان یک استراتژی مهم و حیاتی محسوب شوند.

## اهمیت و ضرورت پژوهش

تجارت الکترونیک شیوه فعالیتهای بازرگانی را تغییر داده است. مدیران بازاریابی از طریق تجزیه و تحلیل درگیری ذهنی مصرف‌کنندگان اطلاعاتی را به دست می‌آورند که موفقیت آنها را در بازار در پی دارد. از این رو بررسی درگیری ذهنی مصرف‌کننده اهمیت فراوانی برای شرکت‌ها در راه رسیدن به اهدافشان دارد، شرکتها برای ادامه حیات خود مجبورند تغییر در درگیری ذهنی مصرف کننده را به خوبی درک نمایند تا بتوانند با موقعیت محیط، سازگار شده و موفقیت خود را تضمین کنند. امروزه پیشرفت اینترنت، جمعیت خرید آنلاین را افزایش داده است. با افزایش تعداد کاربران اینترنت، نحوه استفاده کاربران از این ابزار تعاملی به عنوان بخش موثر در تصمیمات و اقدامات خرید، توجه پژوهشگران و صاحب‌نظران را به خود جلب کرده است، از سوی دیگر عواملی همچون ویژگی‌های وب سایت تمایل افراد را به خرید اینترنتی تحت تاثیر قرار می‌دهد. از این رو توجه به درگیری ذهنی افراد و وب سایت می‌تواند بر روی خرید اینترنتی اثر بگذارد. از آنجا که خرید و فروش در سراسر دنیا به صورت الکترونیکی انجام می‌پذیرد، کشور ما نیز در آینده‌ای نزدیک، ناگزیر از پذیرش و بکارگیری آن خواهد بود.

## اهداف پژوهش

هدف این پژوهش تعیین تاثیر ویژگی‌های وب سایت و درگیری ذهنی بر خرید آنی اینترنتی شهروندان تبریزی که حداقل یک بار از طریق اینترنت پوشاک خریداری کرده‌اند، می‌باشد و در این راستا اهداف فرعی پژوهش به شرح ذیل می‌باشند:

۱-تعیین تاثیر ویژگی‌های لذت گرایانه وب سایت بر خرید آنی اینترنتی

۲-تعیین تاثیر ویژگی‌های نفع گرایانه وب سایت بر خرید آنی اینترنتی

۳- تعیین تاثیر درگیری ذهنی مشتری بر خرید آنی اینترنتی

## فرضیات پژوهش

براساس اهداف پژوهش، فرضیه‌های این پژوهش به شرح زیر می باشند.

۱- ویژگی‌های لذت گرایانه وب سایت تاثیر مستقیم بر خرید آنی اینترنتی دارد.

۲- ویژگی‌های نفع گرایانه وب سایت تاثیر مستقیم بر خرید آنی اینترنتی دارد.

۳- درگیری ذهنی مشتری تاثیر مستقیم بر خرید آنی اینترنتی دارد.

## متغیرهای پژوهش

**متغیرهای مستقل:** ویژگی‌های وبسایت (ویژگی‌های لذت گرایانه، ویژگی‌های نفع گرایانه)، درگیری ذهنی

**متغیر وابسته:** خرید آنی اینترنتی

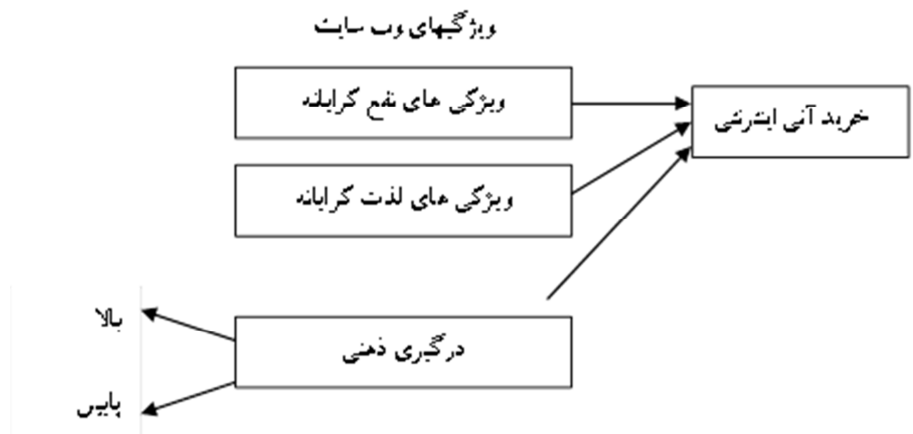

**شکل ۱: مدل مفهومی پژوهش**

## تعاریف و مفاهیم

وب سایت: سایت به مجموعه ای از صفحات وب متصل به هم تلقی می‌گردد که معمولاً روی یک سرور قراردارند و اغلب شامل یک صفحه – home page – است که به عنوان مجموعه ای از اطلاعات توسط یک فرد، گروه یا سازمان تهیه و نگهداری می‌شود (میرزاده، ۱۳۹۴).

درگیری ذهنی: درگیری ذهنی به عنوان فرآیندی که از طریق آن نگرش های افراد در طول تعاملات برانگیخته می‌شود و آن وقت این نگرش ها رفتار افراد را هدایت می کند (شریف و کانتریل، ۱۹۴۷).

خرید آنی اینترنتی: روک[1] خرید آنی اینترنتی را اینگونه تعریف می‌کند، «خرید آنی هنگامی رخ می‌دهد، که مصرف کننده یک میل مزمن اغلب قدرتمند، ناگهانی برای خرید بی‌درنگ چیزی را تجربه می نماید».

## دیدگاهها و مبانی نظری

## ویژگیهای وب سایت

---

[1] Rook, D. W.

سایت به مجموعه ای از صفحات وب متصل به هم تلقی می‌گردد که معمولا روی یک سرور قرار دارند و اغلب شامل یک صفحه اصلی – Home Page – است که به عنوان مجموعه ای از اطلاعات توسط یک فرد، گروه یا سازمان تهیه و نگهداری می‌شود (میرزاده، ۱۳۹۴). ایده اولیه در مورد وب سایت به سال ۱۹۸۰ برمی‌گردد. زمانی که در شهر سرن سوئیس تیم برنرزلی[1] شبکه ENQUIRE را ساخت و همنام کتابی بود که وی از جوانی خود به یاد داشت، اگرچه آنچه وی با ساخت با وب امروزی تفاوتهای زیادی دارد اما ایده اصلی در آن گنجانده شده است. در مارس ۱۹۸۹، برنرزلی یک پیشنهاد را نوشت که به ENQUIRE اشاره داشت و یک سیستم اطلاعاتی پیشرفته را توصیف می کرد. وی با کمک رابرت کایلا[2]، پیشنهاد طراحی تور جهان گستر را در سال ۱۹۹۰ ارائه کرد. اولین مرور وب جهان توسط برنرزلی با عنوان NEXT مورد استفاده قرار گرفت و وی اولین مرورگر وب و تور جهان گستر را در سال ۱۹۹۰ طراحی کرد(برنرزلی، ۱۹۹۰).

ویژگی‌های وب سایت را تحت عناصری از وب سایت که برای مشتریان قابل مشاهده بوده و از فرآیند خرید، پشتیبانی می کنند تعریف می کنیم به عبارتی ویژگی‌های وب سایت عهده دار همان نقشی است که محیط فیزیکی برای فروشگاه ایفا می‌کند.

فروشگاههای فیزیکی از طریق محرکهای فیزیکی و سمعی و بصری در محیط خرید و با مهیا کردن محیطی خوشایند می‌توانند تجربه و انگیزه خرید افراد را ارتقاء دهند، اما خرید اینترنتی که در چارچوب تکنولوژی و محیط مجازی محدود شده است. به خوبی قادر نیست که محرکهای احساسی متنوع و تعامل فیزیکی میان طرفین را ارائه کند. بنابراین در مقایسه با محیط خرید فیزیکی لذت کمتری از طریق فرآیند خرید اینترنتی ایجاد می‌شود. تمایل خریدارانی که به انگیزه لذت جویانه اهمیت می دهند ترجیح می دهند که محصولات موردنیاز خود را از مغازه های فیزیکی و نه از کانال اینترنتی خریداری کنند. از سوی دیگر مصرف کنندگانی که در حال جستجو و خرید از طریق اینترنت هستند تمایل دارند که کارآمد باشند و با اهداف فایده طلبانه برانگیخته می‌شوند. (اسماعیل‌پور، گلدوزیان، ۱۳۹۰). همانند خرید از فروشگاههای فیزیکی، خرید آنلاین نیز تحت تاثیر انگیزه های فایده طلبانه و لذت جویانه است و انگیزه فایده طلبانه محرک قوی تری در خرید آنلاین افراد است و تاثیر قوی تری بر قصد جستجو و قصد خرید آنلاین دارد. در حالی که انگیزه لذت جویانه تاثیر مستقیم بر قصد خرید دارد و از طریق تاثیرگذاری بر قصد جستجو به قصد خرید منجر می‌شود (تو و همکاران[3]، ۲۰۰۷).

در طبقه بندی وب سایت دو انگیزه وجود دارد:

۱- انگیزه فایده طلبانه: در برگیرنده ارزشهایی است که خرید آنلاین را به عنوان یک عمل هدفمند و منطقی و معقول و کارآمد نشان می‌دهد. از دیدگاه فایده طلبانه، انگیزه خرید صرفاً دستیابی کارآمد به یک کالا است. خرید کالاها و خدمات مورد نیاز مصرف کنندگان از طریق اینترنت مزیت های متعددی از قبیل دسترسی آسان و سریع به

---
[1] Tim Berners - Lee
[2] Robertkayla
[3] To et al

اطلاعات، راحتی کار و نبود محدودیت زمانی و مکانی، تنوع بیشتر محصولات، صرفه جویی در هزینه‌ها، دسترسی به کالا و خدمات سفارشی و غیره که اینگونه مزایا می‌تواند معرف این ارزشها باشند (تو، ۲۰۰۷).

۲- انگیزه لذت جویانه: شامل ارزشهایی است که به جنبه احساسی و روانی خرید آنلاین مربوط می‌شود. انگیزه لذت گرایی به ارزش‌های اجتماعی و احساسی که طی فرآیند خرید محصول می‌شود اطلاق می‌گردد (تو، ۲۰۰۷).

## درگیری ذهنی

یکی از موضوعات مهم در رفتار مصرف کننده پردازش اطلاعات توسط مصرف کننده می‌باشد. پردازش اطلاعات توسط مصرف کننده یعنی فرآیندی که از طریق آن مصرف کنندگان با اطلاعات مواجه می‌شوند، به آن توجه می کنند، آن را می فهمند، در ذهن خود حفظ می کنند و سپس آن را برای استفاده آتی بازیابی می کنند. یکی از مشکلات که بارها گزارش شده و بازاریابان با آن برخورد می کنند وادار کردن مصرف کنندگان به دریافت، درک و بخاطر آوردن اطلاعات مربوط به یک کالا یا خدمت می‌باشد. یکی از عوامل مهم در پردازش اطلاعات، درگیری ذهنی مصرف کننده است (موون، مینور، ۱۳۸۸). این مفهوم توجه زیادی را در دو دهه‌ی گذشته به خود جلب کرده است. درگیری ذهنی مصرف کننده به عنوان اهمیت شخصی درک شده، یا علاقه‌ی مرتبط با اکتساب، مصرف و کنارگذاری کالا، خدمت یا ایده تعریف می‌شود (موون، مینور، ۱۳۸۸). درگیری ذهنی از روانشناسی اجتماعی نشات گرفته و برداشتی از درگیری ذهنی است که به روابط بین یک فرد، یک هدف یا موضوع اشاره دارد. برخی پژوهش در مورد درگیری ذهنی را به مطالعات الپورت[1] در سال (۱۹۴۳) پیوند می زنند. او بیان کرد که درگیری ذهنی از رفتارهای بنیادین بوده و از درگیری ضمیر نشأت می گیرد. اما کروگمن درگیری ذهنی را در بازاریابی متداول کرد (حیدرزاده و نوروزی، ۱۳۸۹). اگرچه در رشته بازاریابی تعریف دقیقی از درگیری ذهنی وجود ندارد، اما یک همفکری درباره‌ی آن وجود دارد که درگیری ذهنی، یک سطح فردی و متغیر درونی است که به اهمیت و وابستگی شخصی با اهداف و یا رخدادها بر می‌گردد (عبدالوند و نیک فر، ۱۳۹۰).

پژوهشگران دو نوع مختلف درگیری ذهنی را شناسایی کرده‌اند:

الف) موقعیتی: که در یک دوره زمانی کوتاه رخ می‌دهد، با موقعیت فردی مانند نیاز به تعویض یک کالای خراب، (مثلا یک اتومبیل) مرتبط است. (موون، مینور، ۱۳۸۸). درگیری ذهنی موقعیتی احتمالاً موقتی بوده و هنگامی که خرید کامل می‌شود، ناپدید می‌گردد (دولاکیا[2]، ۲۰۰۱).

ب) پایدار (باثبات): علاقه مداوم به یک محصول یا خدمت اشاره دارد، و بیانگر تعهدی طولانی تر و مرتبط با طبقه کالاست (موون، مینور، ۱۳۸۸).

درگیری ذهنی از دیدگاه پژوهشگران به دو دسته‌ی زیر طبقه بندی می‌شوند:

الف) درگیر ذهنی پایین: محصولاتی که بارها با حداقل تفکر و تلاش خریداری می‌شوند، چرا که حیاتی نیستند و تاثیر زیادی در سبک زندگی خریدار ندارند و مشتری قبلاً آشنایی خوبی در مورد محصول دارد.

---

[1] Allport
[2] Dholakia

ب) درگیری ذهنی بالا: محصولاتی که تنها بعد از توجه دقیق خریداری می‌شوند و خرید انگیزشی است برای مثال، کالاهایی که ارزش سرمایه داری دارند و مشتریان به طور کلی به انجام پژوهش اولیه برای خرید اولیه شان نیاز دارند تا درک صحیحی در مورد محصول بدست آورند.

براساس پژوهشات انجام شده در یک تقسیم بندی کلی، درگیری ذهنی را می‌توان به سه دسته ی کلی تقسیم کرد:

الف) درگیری ذهنی محصول: درگیری ذهنی محصول را می‌توان به عنوان «علاقه یا اهمیت ادراک شده نسبت به طبقه خاصی از محصول » درنظر گرفت.

ب) درگیری ذهنی خرید: درجه توجه که در جهت خرید یک طبقه فعالیت صورت می گیرد را درگیری ذهنی خرید گویند.

ج) درگیری ذهنی تبلیغات: علاقه مصرف کننده برای درگیری در ارتباطات بازاریابی را درگیری ذهنی تبلیغات گویند.

در تقسیم بندی دیگر، ویس[1] (۲۰۰۹) چهار نوع درگیری ذهنی را مشخص کرده است که دو نوع آن نسبت به موارد قبلی جدید می‌باشد:

۱- درگیری ذهنی موقعیتی ۲- درگیری ذهنی پایدار ۳- درگیری ذهنی هنجاری ۴- درگیری ذهنی ریسک

درگیری ذهنی هنجاری به اهمیت یک طبقه محصول برای ارزش ها و احساسات و من مصرف کننده مربوط می‌شود و چهارمین درگیری ذهنی، درگیری ذهنی ریسک می‌باشد. درگیری ذهنی ریسک، ارزیابی اهمیت یا سود آوری ریسک ناشی از خرید یا مصرف یک محصول می‌باشد.

خانم زایکوسکی[2] در مقاله ای در سال ۱۹۸۶، درگیری ذهنی را به سه عامل طبقه بندی کرده است:

۱) عوامل انگیزشی: مربوط به ویژگی‌های فیزیکی محرک از قبیل تفکیک گزینه ها، منبع ارتباط و محتوای ارتباط است.

۲) عوامل موقعیتی: همچنین بر روی سطح درگیری تاثیرگذار است. درگیری موقعیتی باعث بالا بردن علاقه موقتی نسبت به محصول می‌شود. مصرف کنندگانی که در یک موقعیت خاص، بسیار درگیر هستند، ممکن است در مقایسه با آنهایی که با موقعیت کمتر درگیر هستند و در رفتارهای متفاوتی درگیر شوند.

۳) عوامل فردی: شامل نیازهای ذاتی فردی، اهمیت، علاقه و ارزش‌های مربوط به یک شی خاص است.

## خرید آنی اینترنتی

در دهه های اخیر فرآیند تصمیم گیری برای خرید، به طور گسترده ای مورد مطالعه و بررسی قرار گرفته است. فرضیه اصلی در بدنه این دانش این است که انتخابهای مصرف کننده را می‌توان از دیدگاه عقلایی توصیف کرد و این در حالی است که یک انتخاب، پس از بررسی دقیق جنبه های متفاوت کالا و بررسی جایگزین‌های مختلف آن انجام می‌شود (ورسکی و کانمن، ۱۹۷۴). در حالی که در برخی موارد مصرف کنندگان از این حدود عقلایی خارج می‌شوند، در چنین مواردی انتخابها بدون بررسی دقیق در مورد جایگزینهای موجود انجام می‌شوند و اطلاعات کافی در رابطه با

---

[1] weiss
[2] Zaichkowski

کالایی که مشتریان به آن علاقه دارند، در دست نیست و همچنین قصدی قبلی برای خرید آن کالا وجود نداشته است (ورسکی و کانمن، 1981) در چنین لحظاتی است که خریدهای آنی در اینترنت اتفاق می‌افتد.

بسیاری از پژوهش‌ها در مورد خرید آنی به صورت مستقیم یا غیرمستقیم به علل یا سوابق خرید ناگهانی توجه کرده‌اند، متغیرهایی را که موجب این نوع از خرید می‌شوند، می‌توان به دو دسته عوامل و محرک‌های درونی و خارجی تقسیم کرد. در خرید آنی، محرک‌های خارجی اغلب نقش محوری دارند (روک و فیشر، 1995). از این رو هرچه فرد بیشتر در معرض محرک‌های خارجی معین قرار گیرد احتمال انجام خرید آنی بیشتر است (لایر، 1989). علاوه بر این عوامل خارجی، یک سری ویژگی‌های درونی در افراد وجود دارد که در هدایت آنها برای انجام خرید دخیل اند. این عوامل هم شامل درگیری ذهنی می‌شود که درجه تمایل به خرید آنی فرد را تعیین می‌کند و هم شامل محرک‌های درونی از قبیل وضعیت عاطفی و شناختی، ارزیابی هنجاری از خرید آنی و ویژگی‌های جمعیت شناختی می‌شود (داسون و کیم، 2009).

محققان تعاریف ادراکی متعددی از خرید آنی ارائه کرده‌اند: پایرون[1] (1991) تعاریف پیشین را بازبینی کرده و نتیجه گیری می‌کند که هیچ یک از آنها به طور کامل این پدیده جالب و پیچیده را تعریف نمی‌کنند، او سیزده بُعد از تعاریف محققان دیگر را که در مورد خرید آنی متداول بود تعیین کرد، سپس این ابعاد را یکپارچه کرده و تعریفی کوتاه از خرید آنی بدین گونه ارائه داد"خرید آنی نوعی خرید برنامه ریزی نشده است، نتیجه رویارویی با محرکهاست و در محل در موردش تصمیم گیری شده است و پس از خرید ، خریدار واکنش های احساسی و یا شناختی را تجربه می‌کند". روک[2] (1987) وقوع یک خرید ناگهانی را اینگونه تعریف می‌کند: وقتی یک مصرف کننده تمایلی ناگهانی، غالباً قوی و پایدار بر خرید آنی یک کالا پیدا می‌کند، این انگیزه ناگهانی، پیچیده است و ممکن است برای او تعارض احساسی ایجاد نماید.

**پیشینه پژوهش**

نتایج حاصل از پژوهش چوبتراش (1392) در خصوص درگیری ذهنی مصرف کننده و تصمیم گیری خرید حاکی از وجود رابطه معنادار بین درگیری ذهنی مصرف‌کننده و تصمیم‌گیری خرید است.

حسینی (1391) پژوهشی با موضوع تبیین ارزیابی تمایل و وفاداری مشتریان بالقوه به خرید از فروشگاه‌های اینترنتی انجام داده و به این نتیجه رسیده است که فایده ادراک شده بر افزایش تمایل مشتریان به تکمیل فرآیند خرید و تکرار رفتار خرید در فروشگاه‌های اینترنتی تاثیر مستقیم نمی‌گذارد، در واقع افراد از تجربه خرید دیگران تاثیر می‌گیرند، اگر دوستان، خانواده و آشنایان وب سایتی را مفید برای خرید ببینند از آن به خوبی یاد خواهند کرد و به نوعی تبلیغات دهان به دهان برای آن وب سایت به راه خواهند انداخت که این مسئله خود دلیلی برای رجوع و افزایش تمایل به تکرار خرید در وب سایت خواهد شد و همینطور لذت ادراک شده وب سایت بر افزایش تمایل مشتریان به تکمیل فرآیند خرید و تکرار رفتار خرید در فروشگاه‌های اینترنتی تاثیر مستقیم ندارد. لذت بردن از خرید جذابیت های دیداری درک شده و ارزش سرگرمی درک شده وب سایت اثر مثبت غیرمستقیم بر لذت درک شده و تمایل به ادامه خرید دارد، اگر فرد

---
[1] Piron
[2] Rook

خرید کردن در وب سایت را لذت بخش قلمداد کند، با این شرط که به آن اعتماد کند احتمالاً دوباره به آن رجوع کرده و از آن خرید خواهد کرد.

بغدادی و نظری(۱۳۹۲)- شناسایی و بررسی عوامل تاثیرگذار بر خرید آنی آنلاین در فروشگاههای تخفیف گروهی در ایران – به این نتایج دست یافتند که ۱- تمرکز بر بالابودن احساس رضایت از خرید برای مشتریان آنلاین، خود به طرق مختلفی از جمله ایجاد هیجان، سهولت خرید می‌تواند لذت خرید را افزایش دهد و همچنین منجر به حس شعف و خوشحالی ناشی از حضور در فروشگاه اینترنتی و انجام خرید آنلاین شود به گونه‌ای که افراد تمایل به گذراندن بخشی از اوقات فراغت خود در فروشگاههای اینترنتی را داشته باشند و از این کار خود لذت ببرند ۲- تمرکز بر بالا بودن حس تمایل افراد به خرید آنی، از طریق حفظ رضایت ناشی از اینگونه خریدها و کاهش ندامت پس از خرید و تمرکز بر ویژگی‌هایی از خود قانون گذاری افراد را کم رنگ و خصلت آنی بودن را پررنگ می‌کند. ۳- تمرکز بر افزایش تعداد خریدهای آنلاین افراد با توجه به افزایش سهولت و اعتماد در اینگونه خریدها و فرهنگ سازی برای افزایش مطلوبیت خریدهای آنلاین نسبت به خریدهای سنتی.

بخشی زاده و خلیل رودی و رضائیان اکبرزاده (۱۳۹۴) – تاثیر درگیری ذهنی و ویژگی‌های شخصیتی افراد بر خریدآنی پوشاک- نتایج پژوهش نشان می دهند تمایل به خرید آنی پوشاک متاثر از تمایل کلی به خرید آنی و درگیری ذهنی است و تمایل به خرید آنی پوشاک پیش بینی خوبی برای تصمیم خرید آنی پوشاک است یعنی با افزایش تمایل به خرید آنی پوشاک میزان تصمیم خرید آنی پوشاک نیز افزایش می یابد.

مهدیه و چوبتراش (۱۳۹۲)- درگیری ذهنی مصرف کننده و تصمیم گیری خرید – نتایج تحقیقات ایشان حاکی از این است که علاقه مندی به محصول و ارزش مبتنی بر لذت از عوامل موثر در تصمیم گیری خرید هستند. بعضی از افراد در فرآیند خرید نسبت به دیگران بیشتر درگیر می‌شوند، افرادی که سطح بالاتری از علاقه مندی به محصول را دارا هستند زمان بیشتری را صرف تصمیم گیری درباره محصول و برند می کنند.

اسماعیل پور و گلدوزیان (۱۳۹۰) – عوامل موثر برانگیزه خرید آنلاین مصرف کنندگان – نتایج پژوهش ایشان حاکی از این که خریداران اینترنتی هر دو انگیزه فایده طلبانه و لذت جویانه را دارا هستند و به خریداران از فروشگاههای فیزیکی بی شباهت نیستند. خریداران اینترنتی نه تنها از منافع کسب محصول بهره مندمی‌شوند، بلکه احساس لذت در فرآیند خرید اینترنتی را نیز تجربه می کنند.

حسینی (۱۳۹۱) – تبیین ارزیابی تمایل و وفاداری مشتریان بالقوه به خرید از فروشگاههای اینترنتی- نتایج پژوهش وی حاکی است اعتماد به وب سایت، کیفیت وب سایت و هنجارهای ذهنی فرد مستقیما تمایل به تداوم خرید را تحت تاثیر قرار داده، اما فایده ادراک شده و لذت ادراک شده از خرید به طور غیرمستقیم تمایل به تداوم خرید را تحت تاثیر قرار می دهند.

الفت، خسروانی ، جلالی (۱۳۹۰) – شناسایی عوامل موثر بر خرید اینترنتی و الویت بندی آنها با استفاده از ANP فازی – نتایج پژوهش نشان داد، بین ویژگی‌های کالا، فرآیند خرید، ریسک خرید، ویژگی‌های مشتریان، ادراک مشتریان از خرید با میزان خرید اینترنتی ارتباط وجود دارد.

وانگ[1] (2000) - تاثیر ارزش لذت‌گرایانه بر رفتار مصرف‌کننده - به این نتیجه رسید که افرادی که بر ارزش‌های لذت‌جویانه، تمرکز می‌کنند، مصرف کنندگان مدرنی هستند که تمایل دارند درآمد اضافی خود را به منظور لذت بردن، رضایت و رسیدن به مطلوبیت‌های جدید از مصرف کردن، به مصرف برسانند و افرادی لذت‌طلب هستند و بیشتر به ویژگی‌های لذت‌جویانه وب سایت توجه می‌کنند و اگر به آن لذتی که می‌خواهند برسند تکرار خرید اتفاق می‌افتد.

سن[2] و لرمن[3] (2007) - معاینه به بررسی منفی مصرف کننده در وب - به این نتیجه دست یافتند که در انتخاب و تصمیم‌گیری محصولات فایده باور، مصرف‌کنندگان رویکرد به حداکثر رساندن فایده و عملکرد را دارند و قضاوت آن‌ها براساس فعالیت‌های شناختی، هدف‌گرا و به انجام یک وظیفه ضروری گرایش دارد و همچنین مصرف‌کنندگان به پیامدهای فوری مصرف توجه می‌کنند و هدف از مصرف فایده باور، افزایش فایده برای آنان می‌باشد که اگر این ویژگی (نفع گرایانه) بالا باشد دوباره از آن وب سایت خرید خواهند کرد.

**قلمرو پژوهش**

قلمرو مکانی پژوهش کلیه شهروندان تبریزی که حداقل یک بار از اینترنت پوشاک خریداری کردند. قلمرو موضوعی، خرید آنی اینترنتی و عنوان، بررسی تاثیر ویژگی‌های وب سایت و درگیری ذهنی بر روی خرید آنی اینترنتی می‌باشند. قلمرو زمانی پژوهش نیز محدوده بین تیرماه سال 95 تا اسفند ماه سال 95 می‌باشد.

**روش پژوهش**

پژوهش حاضر با توجه به انواع تقسیم‌بندی موجود، از نظر هدف، کاربردی است و از حیث روش از نوع تحقیقات علّی محسوب می‌شود که برای گردآوری داده‌های مورد نیاز متغیرها از روش پیمایشی استفاده شده است. جامعه آماری پژوهش شامل کلیه شهروندان تبریزی است که حداقل یک بار از اینترنت پوشاک خریداری کرده‌اند. با توجه به نامشخص بودن تعداد افراد و عدم دسترسی محقق به آنها جامعه آماری نامحدود در نظر گرفته شده است. روش نمونه‌گیری این پژوهش تصادفی ساده می‌باشد که برای کاهش خطای بالقوه ناشی از منطقه جغرافیایی نمونه آماری از همه‌ی مناطق شهر تبریز انتخاب شد. با توجه به نامحدود بودن جامعه آماری، حجم نمونه آماری پژوهش براساس فرمول کوکران 384 شهروند به دست آمد. روش مورد استفاده برای گردآوری داده‌های این پژوهش، پرسشنامه می‌باشد. همه متغیرهای پژوهش با استفاده از مقیاس پنج گزینه ای طیف لیکرت اندازه‌گیری شده‌اند، که عدد 1 مبین کاملا مخالفم و عدد 5 به معنی کاملا موافقم است. در این پژوهش برای افزایش روایی پرسشنامه از موارد زیر استفاده شده است:

1- بررسی و مطالعه ادبیات موجود، پرسشنامه‌ها و سوالاتی که در پژوهشهای مشابه مورد استفاده قرار گرفته اند.

2- همچنین در این خصوص با اساتید محترم راهنما و مشاور، مشورتهایی صورت گرفته است. بنابراین آزمون روایی از نوع روایی ظاهری بوده است.

برای سنجش پایایی متغیرهای پژوهش از روش آلفای کرونباخ که یکی از متداولترین روش‌های اندازه‌گیری پایایی است، با استفاده از نرم افزار SPSS استفاده شده است. بدین ترتیب 27 سوال پرسشنامه بین 30 نفر توزیع گردید و پس از

---

[1] Wang , CH
[2] Sen , S
[3] Lerman, D

جمع‌آوری پاسخ‌ها ضرایب پایایی این پرسشنامه‌ها محاسبه و مقدار بدست آمده به دلیل اینکه این پرسشنامه از پایایی کافی ($\alpha > 0/7$) برخوردار بود، از آن برای پژوهش استفاده گردید.

**جدول 1: ضرایب پایایی متغیرهای پژوهش**

| سوالات | ضریب آلفای کرونباخ | پایایی ترکیبی CR | متوسط واریانس تبیین شده AVE |
|---|---|---|---|
| درگیری ذهنی | 0/714 | 0/88 | 0/61 |
| ویژگی‌های وب سایت | 0/910 | 0/89 | 0/53 |
| خرید آنی اینترنتی | 0/775 | 0/83 | 0/55 |

## تجزیه و تحلیل داده‌ها
## آمار توصیفی

از نرم افزار[1] برای استخراج شاخص‌های مرکزی (میانگین) و شاخص‌های پراکندگی (انحراف معیار) و همچنین فراوانی متغیرهای جمعیت شناختی و سایر متغیرهای پژوهش استفاده شده است. همچنین برای اطمینان از نرمال بودن توزیع داده‌ها از آزمون کولموگروف اسمیرنوف[2] موجود در نرم افزار استفاده گردیده است. خروجی نرم‌افزار نشان می‌دهند که متغیرها در داخل سطح مورد پذیرش قرار دارد و داده‌ها از توزیع نرمال برخوردار می‌باشند. نتایج شاخص کشیدگی و چولگی و آزمون کولموگروف اسمیرنوف همه متغیرها در جدول 2 ارائه شده است.

**جدول 2: آمار توصیفی نمونه آماری پژوهش**

| گویه‌ها | تعداد | میانگین | انحراف معیار | چولگی | کشیدگی | Z کولموگروف اسمیرنوف | سطح معنی‌داری |
|---|---|---|---|---|---|---|---|
| درگیری ذهنی | 260 | 3/29 | 0/78 | 0/123 | -0/495 | 1/209 | 0/107 |
| ویژگی‌های نفع گرایانه | 260 | 3/59 | 0/8 | -0/372 | -0/482 | 1/581 | 0/103 |
| ویژگی‌های لذت گرایانه | 260 | 3/5 | 0/9 | -0/108 | -0/699 | 1/497 | 0/203 |
| قصد خرید | 260 | 2/49 | 0/94 | 0/385 | -0/263 | 1/599 | 0/102 |

## ویژگی‌های جمعیت شناختی نمونه آماری

برای شناخت بهتر جامعهٔ آماری پژوهش، هشت متغیر جمعیت شناختی جنسیت، سن، تحصیلات، تعداد خرید اینترنتی، ساعت صرف کرده با اینترنت در هفته و نام سایت در نظر گرفته شدند.

**جدول 3: فراوانی متغیرهای جمعیت شناختی نمونه آماری**

| متغیر | شرح | فراوانی | درصد |
|---|---|---|---|

---

[1] Spss 20
[2] Kolmogorov - smirnov

| | | | |
|---|---|---|---|
| جنسیت | زن | ۲۰۲ | ۷۷/۶۹ |
| | مرد | ۵۸ | ۲۲/۳۱ |
| سن | کمتر از ۲۵ سال | ۱۴۵ | ۵۵/۷۷ |
| | ۲۵ تا ۳۴ سال | ۵۲ | ۲۰ |
| | ۳۵ تا ۴۴ سال | ۲۲ | ۸/۴۶ |
| | ۴۵ تا ۵۴ سال | ۲ | ۱۰/۷۷ |
| | ۵۵ تا ۶۴ سال | ۷ | ۲/۶۹ |
| | ۶۵ سال و بالاتر | ۶ | ۲/۳۱ |
| تحصیلات | زیر دیپلم | ۸ | ۳/۸۰ |
| | دیپلم | ۷ | ۲۷/۳۱ |
| | فوق دیپلم | ۲۱ | ۱۰ |
| | لیسانس | ۱۱۰ | ۴۲/۳۱ |
| | فوق لیسانس و بالاتر | ۴۵ | ۱۷/۳۱ |
| تعداد خرید اینترنتی | یکبار | ۳۴ | ۱۳/۸۰ |
| | دو بار | ۳۳ | ۱۲/۶۹ |
| | سه بار | ۵۴ | ۲۰/۷۷ |
| | چهار بار | ۳۸ | ۱۴/۶۲ |
| | پنج بار و بیشتر | ۱۰۱ | ۳۸/۸۵ |
| ساعت صرف کرده با اینترنت در هفته | کمتر از ۱۰ ساعت | ۱۰۸ | ۴۱/۴۵ |
| | ۱۱ تا ۳۰ ساعت | ۹۸ | ۳۷/۶۹ |
| | ۳۱ تا ۵۰ ساعت | ۲۶ | ۱۰ |
| | ۵۱ تا ۷۰ ساعت | ۱۴ | ۵/۳۸ |
| | ۷۱ تا ۹۰ ساعت | ۶ | ۲/۳۱ |
| | ۹۱ تا ۱۱۰ ساعت | ۶ | ۲/۳۱ |
| | بیشتر از ۱۱۱ ساعت | ۲ | ۰/۷۷ |
| نام سایت | دیجی کالا | ۲۳۲ | ۸۹/۳۲ |
| | مد برتر | ۶ | ۲/۳۱ |
| | دیوار | ۶ | ۲/۳۱ |
| | گوگل | ۶ | ۱/۵۴ |
| | بست مد | ۶ | ۲/۳۱ |
| | اینستا | ۴ | ۲/۳۱ |

در ادامه فراوانی هدف از استفاده از اینترنت و همچنین محصولات و خدمات خریداری‌شده از اینترنت آورده شده است.

**جدول ۴: فراوانی هدف از استفاده و محصول خریداری شده از اینترنت**

| متغیر | شرح | فراوانی |
|---|---|---|
| استفاده از اینترنت برای | خرید | ۱۴۴ |
| | جستجوی اطلاعات | ۱۳۵ |
| | کار با شبکه‌های اجتماعی | ۸۲ |
| | سرگرمی | ۶۶ |
| | چت | ۵۱ |
| | ایمیل | ۴۵ |
| | بانکداری | ۴۰ |
| | وبلاگ‌ها | ۱۷ |
| محصول یا خدمت خریداری شده | لباس | ۲۳۵ |
| | کتاب | ۹۳ |
| | نرم افزار یا آموزش | ۶۰ |
| | خدمات مالی و بانکی | ۶۰ |
| | لوازم خانگی | ۵۵ |
| | لوازم الکتریکی | ۲۷ |
| | CD/DVD | ۲۶ |
| | تجهیزات ورزشی | ۲۵ |
| | لوازم بهداشتی و آرایشی | ۱۷ |
| | خدمات مسافرتی | ۱۳ |

## مدل سازی معادلات ساختاری

مدل‌سازی معادلات ساختاری برای آزمون فرضیه‌های ناشی از مدل نظری تحقیق مورد استفاده قرار می‌گیرد. برای انجام تحلیل مدل‌سازی معادلات ساختاری از رویکرد دو مرحله‌ای استفاده شد.

در مرحله اول (مدل اندازه‌گیری) تحلیل‌ها براساس تعیین روابط علی بین متغیرها (گویه‌ها) و سازه‌های نظری انجام می‌گیرد. بدین منظور، با استفاده از نرم‌افزار آموس ۲۰[1] تحلیل عاملی تائیدی انجام شد. بعد از این مرحله، مسیرها یا روابط علی بین سازه‌ها در مدل ساختاری مشخص شدند (مرحله دوم).

مدل سازی معادلات ساختاری، ضمن بررسی روایی و اعتبار هریک از سازه ها یا متغیرهای پژوهش به صورت جداگانه، به طور همزمان به آزمون کلی برازش مدل و برآورد هریک از پارامترهای مدل می پردازد. نرم افزر مدل سازی معادلات

---

[1] Analysis of moment stractures

ساختاری استفاده شده در این پژوهش آموس ۲۰ می‌باشد که برای شناسایی روابط آماری بین گویه ها، هر یک از عامل ها (متغیرها) و روابط بین متغیرهای مستقل (ویژگی‌های وب سایت، درگیری ذهنی) و وابسته (خرید آنی اینترنتی) استفاده شده است. علاوه بر این، محقق می‌تواند در جهت نشان دادن روابط فرض شده در بین متغیرها به تعیین، برآورد، ارزیابی و ارائه مدل در نمودار علّی مسیر بپردازد.

## شاخصهای آزمون برازش مدل

در مدل سازی معادلات ساختاری مجموعه ای از شاخص های نیکویی برازش وجود دارد که مشخص کننده برازش مدل با داده‌هاست. سه گروه کلی از شاخصهای برازش مدل عبارتند از: شاخصهای برازش مطلق[1]، شاخصهای برازش افزاینده[2]، شاخصهای برازش کاهنده[3]. در این پژوهش نیز از هر گروه از شاخص‌های برازش، آن شاخص‌هایی انتخاب گردید که عموما در تحقیقات بازاریابی از آنها برای برازش مدل استفاده می کنند، شاخص‌های انتخاب شده در جدول ۵ ارائه شده است.

**جدول ۵: شاخصهای برای برازش مدل**

| نام شاخص | سطح قابل قبول |
|---|---|
| شاخص نیکویی برازش[4] (GFI) | بالاتر از ۰/۹ |
| ریشه میانگین مربعات خطای بر آوردی[5] (RMSEA) | کمتر از ۰/۰۸ |
| نیکویی برازش اصلاح شده[6] (AGFI) | بالاتر از ۰/۹ |
| شاخص برازش توکر- لویس[7] (TLI) | بالاتر از ۰/۹ |
| شاخص برازش هنجار شده[8] (NFI) | بالاتر از ۰/۹ |
| شاخص برازش تطبیقی[9] (CFI) | بالاتر از ۰/۹ |

## بررسی همبستگی متغیرهای پژوهش

با توجه به نرمال بودن توزیع داده‌ها، همبستگی متغیرهای پژوهش با استفاده از آزمون پیرسون مورد بررسی قرار گرفت. نتایج ارائه شده در جدول ۵ حاکی از معنی‌داری همبستگی متغیرهای پژوهش با ضریب اطمینان ۹۹ درصد می‌باشند.

**جدول ۶: ضرایب همبستگی پیرسون متغیرهای پژوهش**

| | | ویژگی های لذت گرایانه | ویژگی های نفع گرایانه | درگیری ذهنی | خرید آنی |
|---|---|---|---|---|---|
| ویژگی های لذت گرایانه | Pearson Correlation | 1 | .648** | .743** | .683** |

---

[1] Absolute,F:t Indices
[2] Incre mental Fit Indices
[3] Parsimonious Fit Indices
[4] Goodness – of-fit
[5] Root mean square Error of Appoximation
[6] Adjusted Googness – of - fit
[7] Tuker- Lewis Inden
[8] Normal fit Index
[9] Comparative fit Index

| خرید آنی | درگیری ذهنی | ویژگی های نفع گرایانه | ویژگی های لذت گرایانه | | |
|---|---|---|---|---|---|
| 0.000 | 0.000 | 0.000 | | Sig. (2-tailed) | |
| 370 | 370 | 370 | 370 | N | |
| .539** | .615** | 1 | .648** | Pearson Correlation | ویژگی های نفع گرایانه |
| 0.000 | 0.000 | | 0.000 | Sig. (2-tailed) | |
| 370 | 370 | 370 | 370 | N | |
| .740** | 1 | .615** | .743** | Pearson Correlation | درگیری ذهنی |
| 0.000 | | 0.000 | 0.000 | Sig. (2-tailed) | |
| 370 | 370 | 370 | 370 | N | |
| 1 | .740** | .539** | .683** | Pearson Correlation | خرید آنی |
| | 0.000 | 0.000 | 0.000 | Sig. (2-tailed) | |
| 370 | 370 | 370 | 370 | N | |

## آزمون فرضیات پژوهش با مدل سازی معادلات ساختاری(SEM)

قبل از بکارگیری تحلیل عاملی در تحلیل داده‌ها، بایستی از مناسب بودن این روش و کفایت نمونه مورد بررسی اطمینان حاصل کنیم. با توجه به اینکه مقدار بدست آمده برای شاخص KMO بزرگتر از ۰/۶ (۰/۹۲) و نزدیک ۱ می‌باشد، می‌توان نتیجه گرفت تعداد نمونه مورد بررسی برای انجام تحلیل عاملی کافی می‌باشد. مشاهده می‌شود که سطح معنی‌داری آزمون بارتلت (۰/۰۰۰) کمتر از ۰/۰۵ می‌باشد، پس در نتیجه فرض شناخته شده بودن ماتریس همبستگی رد شده و می‌توان گفت تحلیل عاملی برای شناسایی مدل عاملی مناسب است.

**جدول ۷: آزمون KMO و بارتلت**

| ۰/۹۲ | Kaiser-Meyer-Olkin Measure of Sampling Adequacy | |
|---|---|---|
| ۳۵۱ | درجه آزادی | Bartlett's Test of Sphericity |
| ۰/۰۰۰ | سطح معنی داری | |

## مدل اندازه‌گیری درگیری ذهنی

مدل اندازه‌گیری درگیری ذهنی با استفاده از شش شاخص (گویه) اندازه‌گیری می‌شود. با بررسی باقی‌مانده‌های استاندارد شده و شاخص اصلاح مشخص شد که باقی‌مانده‌های استاندارد شده، گویه های پنجم و ششم (d.z5,d.z6) عدد بالاتر از ±۲/۵۸ را نشان می‌دهد که خارج از سطح مورد پذیرش می‌باشد. با حذف گویه پنجم، برازش مدل بهبود یافت و همه شاخص‌های برازش در محدوده مورد پذیرش قرار گرفت. همچنین بارهای عاملی هر پنج شاخص بیش از ۰/۵ می‌باشد و از لحاظ آماری نیز معنی‌دار می‌باشند.

**جدول ۸: شاخص های برازش مدل اصلی و مدل اصلاح شده متغیر درگیری ذهنی**

| شاخص | $X^2$ | df | p | GFI | AGFI | TLI | NFI | CFI | RMSEA | $X^2/df$ |
|---|---|---|---|---|---|---|---|---|---|---|
| مدل اصلی | ۷۲/۴۳۷ | ۹ | ۰/۰۰۰ | ۰/۹۴ | ۰/۸۵۹ | ۰/۹۱۳ | ۰/۹۴۱ | ۰/۹۴۸ | ۰/۱۳۸ | ۸/۰۴۹ |
| مدل اصلاح شده | ۱۴/۵۹۹ | ۵ | ۰/۰۱۲ | ۰/۹۸۵ | ۰/۹۵۴ | ۰/۹۸۱ | ۰/۹۸۶ | ۰/۹۹ | ۰/۰۷۲ | ۲/۹۲ |

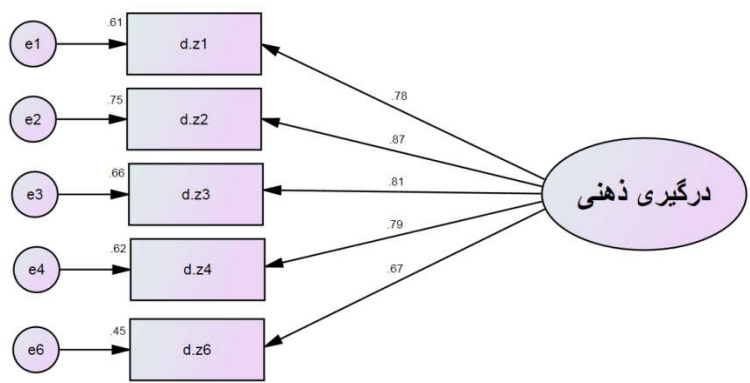

**شکل ۲: نتیجه تحلیل عاملی تائیدی مدل اندازه گیری متغیر درگیری ذهنی**

## مدل اندازه گیری ویژگی های وب سایت

مدل اندازه‌گیری ویژگی‌های وب سایت از نوع تحلیل عاملی مرتبه دوم می‌باشد و از ۱۶ گویه (v.w1 تا v.w16) برای اندازه‌گیری دو عامل ویژگی‌های نفع‌گرایانه و ویژگی‌های لذت‌گرایانه وب سایت استفاده شده است. که از ۱۱ گویه (v.w1 تا v.w11) برای اندازه گیری ویژگی های نفع گرایانه وب سایت و از ۵ گویه دیگر (v.w12 تا v.w16) برای اندازه گیری ویژگی های لذت گرایانه وب سایت استفاده شده است.

**جدول ۹: شاخص های برازش مدل اصلی و اصلاح شده متغیر ویژگی های وب سایت**

| شاخص | $X^2$ | df | p | GFI | AGFI | TLI | NFI | CFI | RMSEA | $X^2/df$ |
|---|---|---|---|---|---|---|---|---|---|---|
| مدل اصلی | ۶۲۶/۲۴۶ | ۱۰۳ | ۰/۰۰۰ | ۰/۸۰۷ | ۰/۷۶۹ | ۰/۷۷۶ | ۰/۷۷۹ | ۰/۸۰۷ | ۰/۱۱۷ | ۶/۰۸ |
| مدل اصلاح شده | ۱۴۵/۶۱۸ | ۴۳ | ۰/۰۰۰ | ۰/۹۳۶ | ۰/۹۰۲ | ۰/۹۱۰ | ۰/۹۰۴ | ۰/۹۳۰ | ۰/۰۸۰ | ۳/۳۸۶ |

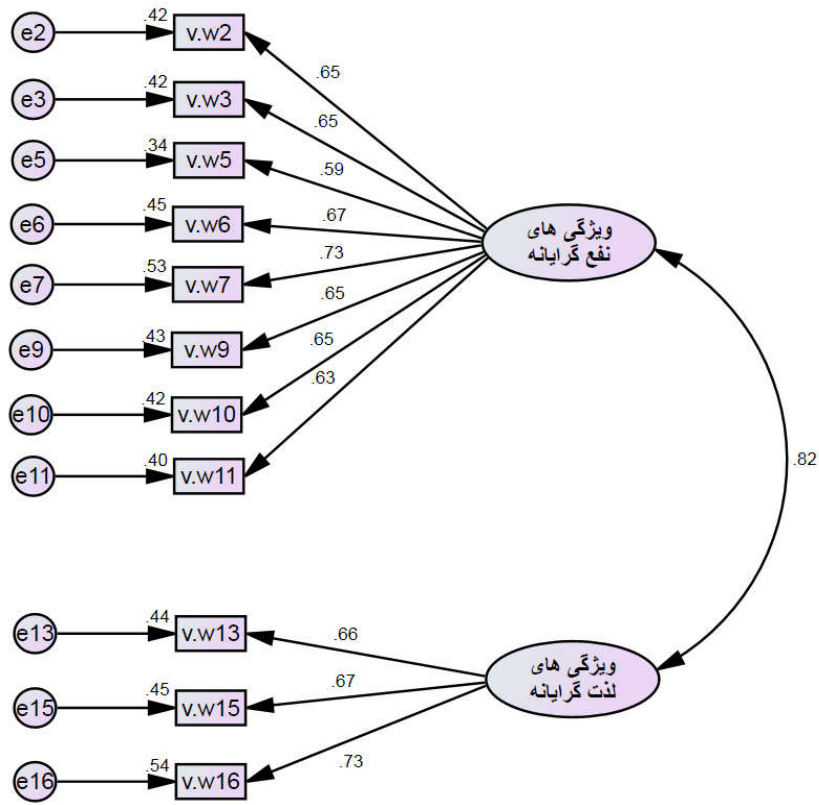

**شکل 3: نتیجه تحلیل عاملی تائیدی مدل اندازه گیری متغیر ویژگی های وب سایت**

## مدل اندازه گیری خرید آنی اینترنتی

برای اندازه گیری مدل تک عاملی خرید آنی اینترنتی از ۵ شاخص (kh.a1 تا kh.a5) که در جدول ۱۰ ارائه شده، استفاده گردید. نتایج تحلیل عاملی نشان می‌دهد که بار عاملی شاخص kh.a5 کمتر از ۰٫۵ می‌باشد. پس از حذف این گویه، مجددا تحلیل عاملی تائیدی اجرا گردید. نتایج نشان می‌دهند که بار عاملی همه گویه‌ها بیشتر از ۰٫۵ می‌باشند و برازش مدل بهبود یافت و همه شاخص‌های برازش در محدوده مورد پذیرش قرار گرفت و از لحاظ آماری نیز معنی‌دار می‌باشند.

**جدول ۱۰: شاخص های برازش مدل اصلی و اصلاح شده متغیر خرید آنی اینترنتی**

| $X^2/df$ | RMSEA | CFI | NFI | TLI | AGFI | GFI | p | df | $X^2$ | شاخص |
|---|---|---|---|---|---|---|---|---|---|---|
| ۶/۲۷۸ | ۰/۱۲ | ۰/۹۵۸ | ۰/۹۵۱ | ۰/۹۱۶ | ۰/۹ | ۰/۹۶۷ | ۰/۰۰۰ | ۵ | ۳۱/۳۸۹ | مدل اصلی |
| ۴/۰۷۵ | ۰/۰۷۱ | ۰/۹۸۹ | ۰/۹۸۵ | ۰/۹۶۶ | ۰/۹۴۳ | ۰/۹۸۹ | ۰/۰۱۷ | ۲ | ۸/۱۵ | مدل اصلاح شده |

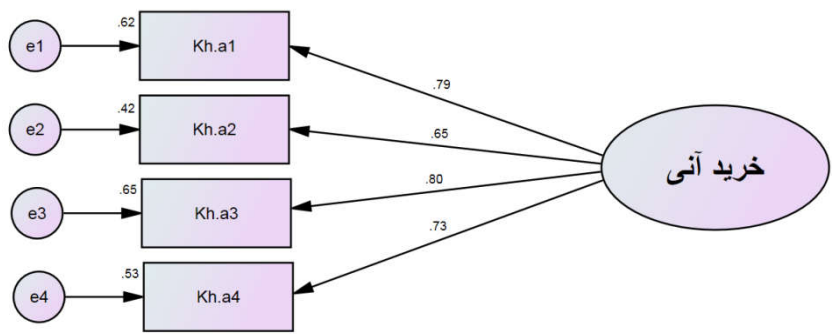

**شکل ۴: نتیجه تحلیل عاملی تائیدی مدل اندازه گیری متغیر خرید آنی اینترنتی**

## مدل ساختاری اصلی

وقتی همه سازه‌ها در مدل اندازه‌گیری (مرحله اول) از روایی لازم برخوردار گردیدند و به برازش رضایت‌بخش رسیدند، آنگاه مدل ساختاری می‌تواند مورد آزمون قرار گیرد و بعنوان مرحله دوم و اصلی تجزیه و تحلیل ارائه گردد. هدف مدل ساختاری تعیین این موضوع است که کدام یک از سازه های پنهان مستقیما یا غیرمستقیم بر مقادیر سایر سازه‌های پنهان در مدل تاثیر می‌گذارد. نتایج آزمون فرضیات پژوهش بر مبنای مدل سازی معادلات ساختاری به ترتیب در جدول ذیل خلاصه شده است.

**جدول ۱۱: آزمون فرضیه ها با استفاده از برآورد ضرایب استاندارد شده**

| فرضیه | مسیرهای فرض شده | ضرایب | خطای معیار | ضرایب استاندارد شده | نسبت بحرانی (t-value) | p | نتیجه |
|---|---|---|---|---|---|---|---|
| H₁ | ویژگی‌های لذت گرایانه وب سایت ← خرید آنی اینترنتی | ۰/۰۵۳ | ۰/۰۴۸ | ۰/۰۵۱ | ۱/۱۱۷ | ۰/۲۶۴ | رد |
| H₂ | ویژگی‌های نفع گرایانه وب سایت ← خرید آنی اینترنتی | ۰/۳۰۴ | ۰/۰۵۹ | ۰/۲۷۶ | ۵/۱۶ | ۰/۰۰۰ | تائید |
| H₃ | درگیری ذهنی ← خرید آنی اینترنتی | ۰/۵۲۳ | ۰/۰۵۴ | ۰/۵۰۴ | ۹/۷۶۳ | ۰/۰۰۰ | تائید |

**جدول ۱۲: شاخص های برازش مدل ساختاری پژوهش**

| x²/df | RMSEA | CFI | NFI | TLI | AGFI | GFI | p | df | X² |
|---|---|---|---|---|---|---|---|---|---|
| ۴/۸۵۹ | ۰/۰۷۳ | ۰/۹۹۶ | ۰/۹۹۶ | ۰/۹۸۹ | ۰/۹۷۵ | ۰/۹۹۶ | ۰/۰۰۷ | ۳ | ۱۴/۵۷۷ |

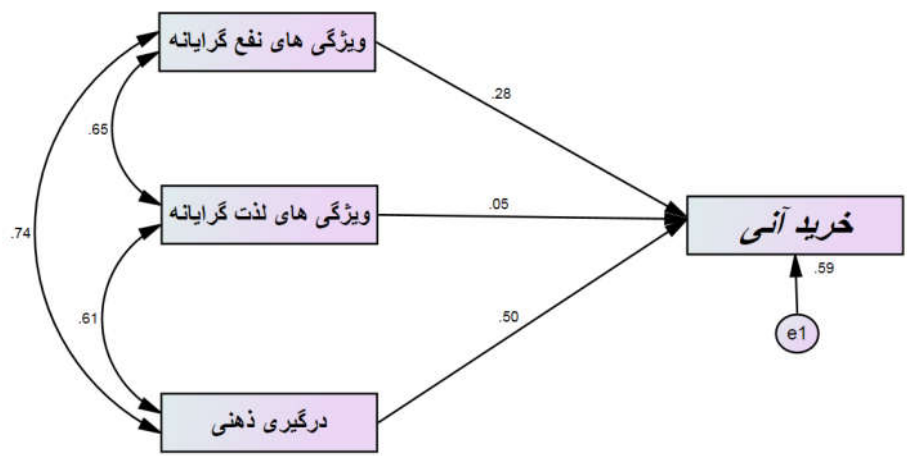

**شکل ۵: نتیجه بدست آمده برای مدل ساختاری**

## نتیجه گیری

نتایج پژوهش در سه بخش به شرح زیر قابل بحث و بررسی می‌باشد:

۱- تاثیر ویژگی های لذت گرایانه وب سایت بر خرید آنی اینترنتی

نتایج آزمون فرضیه $H_1$ نشان داد که ویژگی های لذت گرایانه وب سایت بر خرید آنی اینترنتی تاثیر ندارد. یعنی ویژگی‌های لذت گرایانه در وب سایت تاثیر غیر مستقیمی بر خرید آنی اینترنت در یافته های پژوهش داشته است. از تحلیل این فرضیه چنین استنباط می گردد که مشتریان خرید اینترنتی؛ ویژگی لذت بردن از وب سایت را در جذابیت های دیداری درک شده و ارزش سرگرمی درک شده وب سایت اثر غیرمستقیم بر لذت درک شده و تمایل به ادامه خرید دارد، اگر فرد خرید کردن در وب سایت را لذت بخش قلمداد کند، با این شرط که به آن اعتماد کند احتمالاً دوباره به آن رجوع کرده و از آن خرید خواهد کرد. بنابراین ویژگی لذت گرایانه وب سایت به صورت درک شده برای مشتریان تاثیر غیر مستقیم دارد. از یافته‌های محققان و نتایج حاصل از یافته های پژوهش می توان چنین گفت که ویژگی های لذت گرایانه وب سایت می تواند خود به طرق مختلف از جمله ایجاد هیجان، حس کنجکاوی، سهولت خرید، استفاده طرح های گیرا، رنگ ها، موسیقی و ویدئو های جذاب روی صفحات احساس رضایت خریداران اینترنتی را افزایش داده و لذت خرید توسط کاربران را هم به صورت مستقیم و هم بصورت غیر مستقیم (درک شده) بیشتر کند. و با خلق چنین طرز تفکری در خریداران اینترنتی می توان گفت که ویژگی لذت گرایانه مشتریان در تبلیغ و بازاریابی وب سایت ها به خوبی انجام یافته است.

۲- تاثیر ویژگی های نفع گرایانه وب سایت بر خرید آنی اینترنتی

نتایج آزمون فرضیه $H_2$ نشان داد که ویژگی های نفع گرایانه وب سایت بر خرید آنی اینترنتی تاثیر مستقیمی دارد. یعنی ویژگی‌های نفع گرایانه وب سایت بر خرید آنی اینترنتی تأثیر مستقیم دارد و با توسعه ویژگی‌های نفع گرایانه وب سایت خرید آنی اینترنتی مشتریان افزایش می‌یابد.

می توان گفت که ویژگی های نفع گرایانه وب سایت در برگیرنده ارزشهایی است که خرید آنلاین را به عنوان یک عمل هدفمند ، منطقی و معقول و کارآمد نشان می دهد. یعنی مشتریان هر چقدر هدفمند، منطقی در وب سایت جهت خرید اینترنتی اقدام کنند به همان اندازه نیز در خرید آنی اینترنتی منطقی و معقولانه رفتار خواهند کرد. و هرچه محصولات متنوع تر، قیمت ها ارزان تر، کیفیت محصول بالاتر و جستجوی منابع اطلاعاتی مشتریان فایده طلبانه بیشتر باشد به همان اندازه نیز به نفع خریداران اینترنتی می باشد و در نتیجه چنین وب سایتی از ویژگی های نفع گرایانه برخوردار بوده است و خریداران آنی اینترنتی بیشتر می باشد. و با خلق چنین طرز تفکری در خریداران اینترنتی می توان گفت که ویژگی نفع گرایانه مشتریان در تبلیغ و بازاریابی وب سایت ها به خوبی انجام یافته است.

3- تاثیر درگیری ذهنی بالا و پایین بر خرید آنی اینترنتی

نتایج آزمون فرضیه H3 نشان داد که درگیری ذهنی بالا و پایین بر خرید آنی اینترنتی تاثیر مستقیمی دارد. یعنی با افزایش درگیری ذهنی مشتریان، خرید آنی اینترنتی آنها افزایش می‌یابد و درگیری ذهنی بالا باعث افزایش خرید آنی اینترنتی شده و درگیری ذهنی پائین باعث کاهش خرید آنی اینترنتی مشتریان می‌شود

از تحلیل این فرضیه چنین استنباط می گردد که درگیری های ذهنی بالا و پایین بر خرید آنی اینترنتی تاثیر مستقیمی دارد و درگیری ذهنی می تواند بیشتر بر قیمت محصول ، لذت خرید اینترنتی، علاقه مندی به محصولات و درگیری صرف زمان، درگیری اعتماد به سایت و غیره باشد؛ یعنی مصرف کنندگان با درگیری خرید بالا، قیمت بیشتری پرداخت می کنند و با درگیری خرید پایین قیمت کمتر، همچنین مصرف کنندگان با درگیری بالا، زمان بیشتری را نسبت به درگیری محصول پایین می گذرانند و برعکس.

درگیر ذهنی بر بالا بردن حس تمایل افراد به خرید آنی، از طریق حفظ رضایت ناشی از این گونه خریدها و کاهش ندامت پس از خرید، و تمرکز بر ویژگی هایی که خود قانون گذاری افراد را کم رنگ و خصلت آنی بودن را پر رنگ می کند.

مصرف کنندگان با درگیری ذهنی بالاتر در مورد محصولات و جایگزین های متفاوت آن فعالانه جستجو می کنند. آنها منابع اطلاعاتی متفاوتی را جستجو و تلاش های آگاهانه ای را صرف جمع آوری اطلاعات می کنند، پردازش اطلاعات توسط آنان عمیق تر صورت گرفته و تصمیم گیری خریدشان بسیار پیچیده تر است. ارزیابی پس از خرید این مصرف کنندگان سنجیده تر و آگاهانه تر و راضی و خشنود کردن شان به مراتب دشوارتر است، بازاریان نیازمندند که تلاش بیشتری را صرف رضایت آنها نمایند، چرا که آنان در خرید دیگران به عنوان جستجو کنندگان عقاید و نظرات خریداران اینترنتی موثر باشند.

## پیشنهادهای پژوهش

با توجه به یافته‌های پژوهش، پیشنهاداتی به شرح زیر در جهت ارتقاء ویژگی های بهتر وب سایت و کاهش درگیری های ذهنی شهروندان بر روی خرید اینترنتی ارائه می‌گردد.

1- پیشنهادی کاربردی در جهت تقویت لذت گرایانه وب سایت در کاربران:
- استفاده از طراحی حرفه ای برای وب سایت
- قراردادن اخبار سایت و رسانه ها در وب سایت

- تماس و ارتباط روشن با مشتری و دسترسی آسان به منوها (استفاده راحت مشتریان از سایت)
- حفظ حریم شخصی کاربر
- قرار دادن لوگوی بانک ها در وب سایت
- تعدد راه های پرداخت
- ریسپانسیو بودن سایت
- استفاده از تکنولوژی مدرن و جدید

2- پیشنهادهای کاربردی به منظور درگیری های ذهنی کاربران
- برقراری ارتباط بین افرادی با قصد خرید الکترونیکی از سایت با سایر مشتریانی که قبلاً از سایت فروشگاه خرید کرده اند و استفاده از نظرات آنان؛
- امکان ایجاد دوستی بین مشتریان فروشگاه از طریق شبکه های اجتماعی

3- پیشنهادی کاربردی به منظور توسعه نفع گرایانه مشتریان
- قرار دادن طیف گسترده‌ای از محصولات و امکان فروش آنها در سایت
- اطلاع رسانی کامل در مورد کالاها و ذکر مواردی مانند: قیمت کالا، هزینه ارسال کالا و یا زمان تحویل کالا.
- ارائه قیمتی کمتر از فروشگاه های سنتی

امکان بازگشت کالای معیوب و تعویض جنس خریداری شده، تلاش در جهت رفع مشکلات احتمالی و خدمات پس از فروش مناسب.

## منابع و مآخذ: